\documentclass[12pt]{article}
\usepackage{graphicx} 
\usepackage{hyperref}
\usepackage{url}
\usepackage{listings,jvlisting}
\usepackage{amssymb}
\usepackage{tabularx}
\usepackage{amsmath}
\usepackage{mdframed}
\usepackage{graphicx}
\usepackage{setspace}
\usepackage{here}
\usepackage{authblk}
\usepackage{mathtools}
\usepackage{amsthm}
\usepackage{algorithmic}
\usepackage{algorithm}
\usepackage{cite}

\def\al{{\alpha}}
\def\be{{\beta}}

\def\ep{{\varepsilon}}

\def\si{{\sigma}}

\def\bbe{{\text{\boldmath $\beta$}}}

\def\bth{{\text{\boldmath $\theta$}}}

\def\Sh{{\hat S}}

\def\bbeh{{\hat \bbe}}
\def\beh{{\hat \be}}

\def\sih{{\hat \si}}

\def\bet{{\tilde \be}}

\def\bbeh{{\widehat \bbe}}

\def\De{{\Delta}}

\def\Deh{{\widehat \De}}

\def\x{{\text{\boldmath $x$}}}
\def\y{{\text{\boldmath $y$}}}

\def\M{{\text{\boldmath $M$}}}

\def\V{{\text{\boldmath $V$}}}

\def\X{{\text{\boldmath $X$}}}

\def\behb{{\overline \beh}}
\def\behbb{{\overline \behb}}

\def\xo{{\overline x}}
\def\yo{{\overline y}}

\def\mh{{\hat m}}

\def\Vh{{\widehat V}}

\def\Nc{{\cal N}}

\begin{document}
\title{Random Effect Restricted Mean Survival Time Model}
\author[1]{Keisuke Hanada}
\author[1]{Masahiro Kojima\footnote{Address: Biometrics Department, R\&D Division, Kyowa Kirin Co., Ltd.
Otemachi Financial City Grand Cube, 1-9-2 Otemachi, Chiyoda-ku, Tokyo 100-004, Japan. Tel: +81-3-5205-7200 \quad
E-Mail: masahiro.kojima.tk@kyowakirin.com}}
\affil[1]{Kyowa Kirin Co., Ltd}

\maketitle
\abstract{\noindent
The restricted mean survival time (RMST) model has been garnering attention as a way to provide a clinically intuitive measure: the mean survival time. RMST models, which use methods based on pseudo time-to-event values and inverse probability censoring weighting, can adjust covariates. However, no approach has yet been introduced that considers random effects for clusters. In this paper, we propose a new random-effect RMST. We present two methods of analysis that consider variable effects by i) using a generalized mixed model with pseudo-values and ii) integrating the estimated results from the inverse probability censoring weighting estimating equations for each cluster. We evaluate our proposed methods through computer simulations. In addition, we analyze the effect of a mother's age at birth on under-five deaths in India using states as clusters.
}
\par\vspace{4mm}

\section{Introduction}
A restricted mean survival time (RMST) is statistically robust and independent under the assumption of the proportional hazards property, and it gives clinically interpretable results in terms of mean survival time~\cite{uno2014moving}. When comparing mean survival times between groups, covariates must be adjusted to account for bias in the baseline information. To analyze time-to-event data, how to handle right-censored data should be considered. Unlike Cox regression, RMST cannot handle a partial likelihood over uncensored data. Andersen et al.~\cite{andersen2004regression} proposed a covariate adjustment method that imputes the pseudo-value (PV) derived from the leave-one-out method for censored data and uses the generalized linear model to analyze completed data. A feature of the PV method is that it can also use censored data. Tian et al.~\cite{tian2014predicting} proposed a method using inverse probability censoring weighting (IPCW) to account for censored data. There is no approach that considers random effects for clusters.

Considering a random effect in the model allows for different effect sizes among the clusters, while enabling a common overall effect size. In practice, treatment effects may differ across clusters, for example, by country or populations classified by certain biomarker values. Analysis can be performed to account for heterogeneity across clusters. By estimating the parameters of the distribution that the variant effect follows, we can also check how much heterogeneity there may be. In addition, the amount of effect within each cluster can be accurately estimated according to the degree of heterogeneity using reduced empirical Bayesian estimation.

In this paper, we propose the random-effect RMST model. We present two methods of analysis that take into account variable effects by i) using a generalized mixed model with PVs and ii) integrating the estimated results from the IPCW estimating equations for each cluster. We evaluate our proposed methods through computer simulations. In addition, using states as clusters, we analyze the effect of a mother's age at birth on under-five deaths in India.

This paper is organized as follows. Section 2 introduces two already proposed RMST models, proposes the new random-effect RMST model, and introduces the method used to estimate the model parameter and variance parameter of the random effect. Section 3 describes the setting and results of the computer simulations. Section 4 presents the results of an analysis of data from the eight Empowered Action Group (EAG) states in India. Finally, Section 5 discusses the results.

\section{Method}
We assume that the survival time is $T$, the right-censoring variable is $C$, and the $p$ covariates are $\X=(X_1,X_2,\ldots,X_p)^T$. There are $I$ clusters, the number of subjects in the $i$-th cluster is $N_i$, and the total sample size is $N(=\sum^I_{i=1}N_i)$. The restricted time for calculating the RMST is a time point $\tau$ of clinical interest. The observable data of the $n_{ij}$-th subject in cluster $i$ are $Y_{ij}=\min(T_{ij},C_{ij},\tau)$, $\Delta_{ij}=I(T_{ij}\leq \min(C_{ij},\tau))$, and $\X_{ij}$.

We propose two types of RMST model with random effects. The first one is based on a method using PVs proposed by Andersen et al.~\cite{andersen2003generalised}, and we call it the PV method. The other is based on the IPCW estimating equation proposed by Tian et al \cite{tian2014predicting} and two-stage estimation method, and we call this method the IPCW method.
We present the first method using PVs. For censored data, PVs are generated by the leave-one-out method. The RMST estimator for all subjects is $\mh(\tau)=\int^\tau_0 \Sh(t)dt$, where $\Sh(t)$ is the Kaplan-Meier estimator. The RMST estimator excluding subject $n_{ij}$ is $\mh_{-ij}(\tau)$, and the PV of subject $n_{ij}$ is $N\mh(\tau)-(N-1)\mh_{-ij}(\tau)$. In this paper, we compute the PVs within each cluster to account for the heterogeneity among clusters. The RMST estimator in the $i$-th cluster is 
\begin{align}
\mh^{(i)}(\tau)=\int^\tau_0 \Sh^{(i)}(t)dt,
\end{align}
and the PVs are calculated from 
\begin{align}
N_i\mh^{(i)}(\tau)-(N_i-1)\mh^{(i)}_{-ij}(\tau).
\end{align}
Because all censored data are complemented by PVs, the RMST can be calculated by a model analysis without considering the censoring. We consider an analysis that accounts for a random effect using a generalized linear mixed model with the link function as $g(\mu_{ij})=\bbe^T\x_{ij}+v_i$, where $\mu_{ij}=E[Y_{ij}|v_i]$ and $v_i$ is a random effect distributed as a normal distribution of $\Nc(0,\si^2_v)$. The parameters of $\bbe$ and $\si^2_v$ can be estimated by a generalized estimating equation, such as Laplace approximation, adaptive Gaussian-Hermite quadrature, or penalized quasi-likelihood. Note that we introduce the generalized estimating equation method by referring to Jiang and Nguyen\cite{jiang2007linear}. We denote the observed time-to-event values and the generated PVs as $Y$ without distinction. We assume the expectation of $Y$ is 
\begin{align}
E[Y_{ij}]=\int h(\x_{ij}^T\bbe+\si_vu_i)f(u_i)du_i\equiv M_i(\bth),
\end{align}
where $h(\cdot)=g^{-1}(\cdot)$, $u_i$ is a standard normal variable, $f(u_i)$ is a standard normal density function, and $\bth=(\bbe^T,\si_v)^T$. The first derivatives are
\begin{align}
\label{eq:derive}
\frac{\partial M_i(\bth)}{\partial \bth}=
\begin{pmatrix}
\frac{\partial M_i(\bth)}{\partial \bbe} \\
\frac{\partial M_i(\bth)}{\partial \si_v} \\
\end{pmatrix}
=
\begin{pmatrix}
\x_{ij}\int h^\prime(\x_{ij}^T\bbe+\si_vu_i)f(u_i)du_i \\
\int h^\prime(\x_{ij}^T\bbe+\si_vu_i)u_if(u_i)du_i \\
\end{pmatrix}
.
\end{align}
We assume $V[Y_{ij}]\equiv V_0$. The generalized estimating equation is 
\begin{align}
\label{eq:equation}
\sum^I_{i=1}\frac{\partial \M_i(\bth)^T}{\partial \bth}\Vh_i^{-1}(\y_i-\M_i(\bth))=0,
\end{align}
where $\V_i$ is a working covariance matrix for the $i$-th cluster. The initial value of $\V_i$ is an $N_i\times N_i$ identity matrix $\V_i$. Parameters $\bth$ are estimated from Equation (\ref{eq:derive}). Next, we calculate $\V_i$ using the estimated values again and iterate through Equations (\ref{eq:derive}) and (\ref{eq:equation}). If there is no difference between the current estimation values and the previous estimation values, the iterations are stopped, and the last estimation values are output as the generalized estimating equation estimators.\\
From here, we introduce the shrinkage estimator and adjusted confidence interval for each cluster using the empirical Bayes estimator when $Y$ is distributed as a normal distribution~\cite{kubokawa2010corrected}. Specifically, we are interested in the difference between groups. We assume the indicator variable for the group is $x_{1ij}$ and the parameter of the group is $\be_{1}$.
The empirical Bayes estimator of $\be_{1i}$ for the $i$-th cluster is 
\begin{align}
\beh_{1i}^{EB}=\beh_{1i}+\frac{\sih^{2}}{\sih^{2}+\sih_v^{2}}(\beh_1-\beh_{1i}),
\end{align}
where $\sih^2$ is the square of the standard error of $\beh_{1}$, $\xo_{1i}=\frac{1}{N_i}\sum^{N_i}_{j=1}x_{1ij}$, and $\yo_{1i}=\frac{1}{N_i}\sum^{N_i}_{j=1}y_{1ij}$. The adjusted $100(1-\al)\%$ confidence interval is $\left[\beh_{1i}^{EB}-z_{\al/2}\sqrt{\frac{\sih^2\sih_v^2}{\sih^2+\sih_v^2}},\beh_{1i}^{EB}+z_{\al/2}\sqrt{\frac{\sih^2\sih_v^2}{\sih^2+\sih_v^2}}\right]$. If $Y$ is distributed as a log-normal distribution, then we refer the reader to Berg and Chandra \cite{berg2014small}. For the log-normal distribution, the formula for the error in the confidence interval is very complicated.

The second proposed method is based on the IPCW estimating equation by Tian et al. \cite{tian2014predicting}. Because the IPCW estimating equation is unable to account for a random effect, we consider a two-stage estimation, first estimating the coefficient parameters by cluster and then accounting for heterogeneity across clusters in the estimated results. For the $i$-th cluster, the parameters are estimated by the following IPCW estimating equation:
\begin{align}
\frac{1}{N_i}\sum^{N_i}_{j=1}\frac{\Deh_{ij}}{KM(y_{ij})}\x_{ij}(y_{ij}-h(\x_{ij}^T\bbe))=0,
\end{align}
where $\Deh_{ij}=I(Y_{ij}\leq C_{ij})$ and $KM(y_{ij})$ is the Kaplan-Meier estimator of the censoring time $C_{ij}$. We assume that the estimator of the parameters in the $i$-th cluster is $\bbeh_i$. We assume that the variable of our particular interest is the $1$-th parameter of the group, and the standard error of $\beh_{1i}$ is $\sih_{1i}$. There is a random effect $v\sim\Nc(0,\si^2_v)$. The estimator of $\si^2_v$ is 
\begin{align}
\sih^2_v=\frac{Q-(I-1)}{c},
\end{align}
where $Q=\sum^I_{i=1}\sih_{1i}^{-2}(\beh_{1i}-\behb_{1})^2$, 
\begin{align}
\behb_{1}=\frac{\sum^I_{i=1}\sih_{1i}^{-2}\beh_{1i}}{\sum^I_{i=1}\sih_{1i}^{-2}} \mbox{, and } c=\sum^I_{i=1}\sih_{1i}^{-2}-\frac{\sum^I_{i=1}\sih_{1i}^{-4}}{\sum^I_{i=1}\sih_{1i}^{-2}}.
\end{align}
Here, $\behb_{1}$ is the pooled mean estimator without a random effect. If $Q<I-1$, then $\sih^2_v=0$. The pooled mean estimator $\behbb_1$ with a random effect is $\sum^I_{i=1}(\sih_{1i}^{2}+\sih_v^{2})^{-1}\beh_{1i}/\sum^I_{i=1}(\sih_{1i}^{2}+\sih_v^{2})^{-1}$. Each shrinkage estimator is $\bet^{EB}_{1i}=\beh_{1i}+\frac{\sih_i^{2}}{\sih_{1i}^{2}+\sih_v^{2}}(\behbb_1-\beh_{1i})$~\cite{quan2013empirical}. The adjusted $100(1-\al)\%$ confidence interval is $\left[\bet^{EB}_{1i}-z_{\al/2}\sqrt{\frac{\sih^2_v\sih^2_{1i}}{\sih^2_v+\sih^2_{1i}}},\bet^{EB}_{1i}+z_{\al/2}\sqrt{\frac{\sih^2_v\sih^2_{1i}}{\sih^2_v+\sih^2_{1i}}}\right]$~\cite{yoshimori2014second}. For the log-normal distribution, we refer the reader to Slud and Maiti (2006)~\cite{slud2006mean}.

\section{Simulation}
We evaluated the performance of our proposal method using Monte Carlo simulation. Data-generating model 1 is
\begin{align}
\label{eq:sim_model1}
Y_{ij}=\exp(\be_0+\be_1 x_{1ij}+v_i)\ep_{ij},
\end{align}
where $\be_0=1$, $\be_1=0.5$, $v_i\sim\Nc(0,\si^2_v)$, $\si^2_v=0.3^2$, and $\ep_{ij}\sim Exp(1)$. Moreover, $x_{1ij}$ is 1 for the treatment group and 0 for the control group. The groups are assigned randomly and uniformly. In addition, we consider a model with covariates. Data-generating model 2 is 
\begin{align}
\label{eq:sim_model2}
Y_{ij}=\exp(\be_0+\be_1 x_{1ij}+\be_2 x_{2ij}+\be_3 x_{3ij}+v_i)\ep_{ij},
\end{align}
where $\be_0=1$, $\be_1=0.5$, $\be_2=0.1$, $\be_3=-0.5$, $v_i\sim\Nc(0,\si^2_v)$, $\si^2_v=0.3^2$, and $\ep_{ij}\sim Exp(1)$. Furthermore, $x_{1ij}$ is 1 for the treatment group and 0 for the control group, $x_{2ij}$ is $\Nc(1,0.5^2)$, and $x_{3ij}$ is set to 1 or 0 randomly. The groups are assigned randomly and uniformly. The censoring probability is set to $0.1$, $0.5$, and $0.9$. We consider the three numbers of clusters: 5, 8, and 10. The number of simulations is 10,000, and the total sample size is 400 or 1,000. The restricted time $\tau$ is 5 months.
The evaluation points are
\begin{enumerate}
    \item Average bias (the estimated result - true value) of $\beta_1$;
    \item Mean squared error (MSE) of $\beta_1$;
    \item Coverage probability of the confidence interval of $\beta_1$;
    \item Length of the confidence interval of $\beta_1$;
    \item Average bias of $\si^2_v$;
    \item MSE of $\si^2_v$.
\end{enumerate}
The reference model for comparing performance is a model that excludes the variable effect $v_i$ from Equations (\ref{eq:sim_model1}) and (\ref{eq:sim_model2}).
The estimator of parameter $\hat{\beta}_1$ from the generating model may not converge to the true value of $\beta_1$. Therefore, we assume that the true parameter value for the treatment group of the RMST model is $\hat{\beta}_1$. Moreover, we calculate the approximate true value of parameter $\bar{\beta}_1$ using the 100,000 data points for each group. The true value $\bar{\beta}_1$ is given in Appendix \ref{ap:sim_beta}.
To evaluate the coverage probability of the confidence interval of $\beta_1$, we calculate the $95\%$ confidence interval as $[\behb_{1} - z_{97.5} \sum_{i=1}^I \hat{\sigma}_i^{-2}, \behb_{1} + z_{97.5} \sum_{i=1}^I \hat{\sigma}_i^{-2}]$.

The simulation results with censoring probabilities of $0.1$ and $0.5$ are presented in Tables \ref{table:model1} and \ref{table:model2}. The result with a censoring probability of $0.9$ is given in Appendix \ref{ap:cnsr0.9}.

\textbf{Average bias of $\beta_1$:}
The bias of the PV method is generally small in all situations, but the bias increases with the number of clusters when the censoring probability is large (0.9) regardless of the model.
The bias of the IPCW method is also small for model 1 and when the censoring probability is small (0.1).
However, under conditions with moderate to large censoring probabilities, the bias of the IPCW method increases significantly more than the bias of the PV method when the number of clusters is increased and the number of subjects is small.

\textbf{MSE of $\beta_1$:}
The MSE of the PV method decreases as the number of subjects is increased for any model and censoring probability.
The MSE decreases under a large censoring probability of 0.9, but the correction due to PVs may have an impact on it.
The MSE of the IPCW method also decreases as the number of subjects is increased, but the MSE of the IPCW method is higher than that of the PV method because of the larger censoring probability.

\textbf{Coverage probability of the confidence interval of $\beta_1$:}
The coverage probability of the PV method approaches a nominal level as the number of subjects increases.
Increasing the number of clusters improves coverage probability if the number of subjects is sufficiently large, but decreases it if the number of subjects is not large enough.
The IPCW method shows a similar trend to the PV method, but the effect of the censoring probability is more significant.
In particular, the coverage probability falls below 0.5 when the number of clusters is increased with a small number of subjects under a large censoring probability of 0.9.

\textbf{Length of the confidence interval of $\beta_1$:}
The length of the confidence interval tends to be the same for both methods.
That is, an increase in the number of subjects shortens the length, but an increase in the number of clusters does not change the length much. Under moderate to large censoring probabilities, the confidence interval of model 2 tends to be longer than that of model 1.

\textbf{Average bias of $\si^2_v$:}
The bias of the PV method decreases slightly as the number of clusters increases, but increases as the number of subjects increases.
The result for the IPCW method is similar to that of the PV method, with bias decreasing with an increase in the number of clusters and increasing with an increase in the number of subjects.
However, the PV method has a positive bias, in contrast to the IPCW method, which has a negative bias.
For medium to large censoring probabilities and a small number of subjects, the bias of the IPCW method increases even if the number of clusters increases.

\textbf{MSE of $\si^2_v$:}
The MSE of the PV method decreases as the number of subjects or clusters are increased.
The MSE of the IPCW method remains smaller than that of the PV method in situations where the number of subjects is large and the censoring probability is small.
However, when the censoring probability is moderate to large, the MSE of the IPCW method increases as the number of clusters increases.

\section{Actual data analysis}
\begin{figure}[H]
  \begin{center}
  \includegraphics[width=15cm]{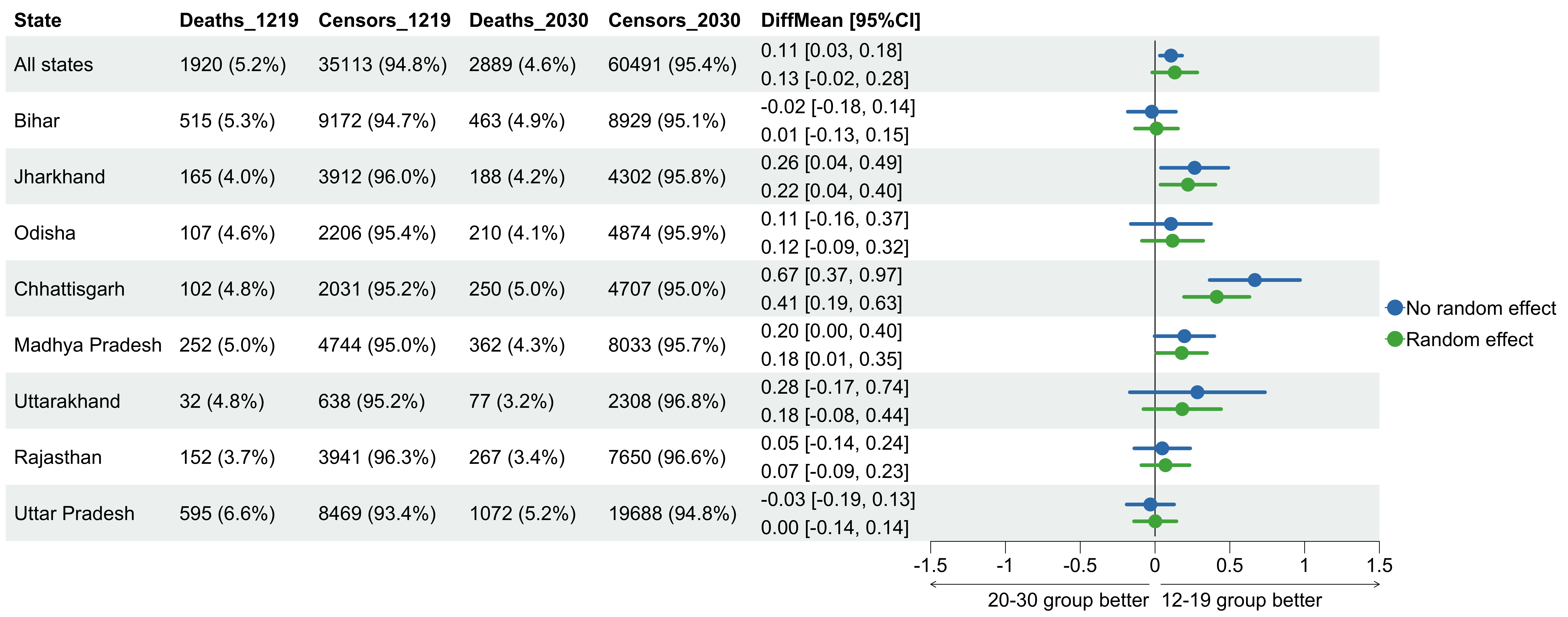}
  \caption{(PV method) Forest plot of the difference between the restricted mean survival times (12--19 group vs 20--30 group)}
  \label{Andersen1}
        \footnotesize{}
  \end{center}
\end{figure}

\begin{figure}[H]
  \begin{center}
  \includegraphics[width=15cm]{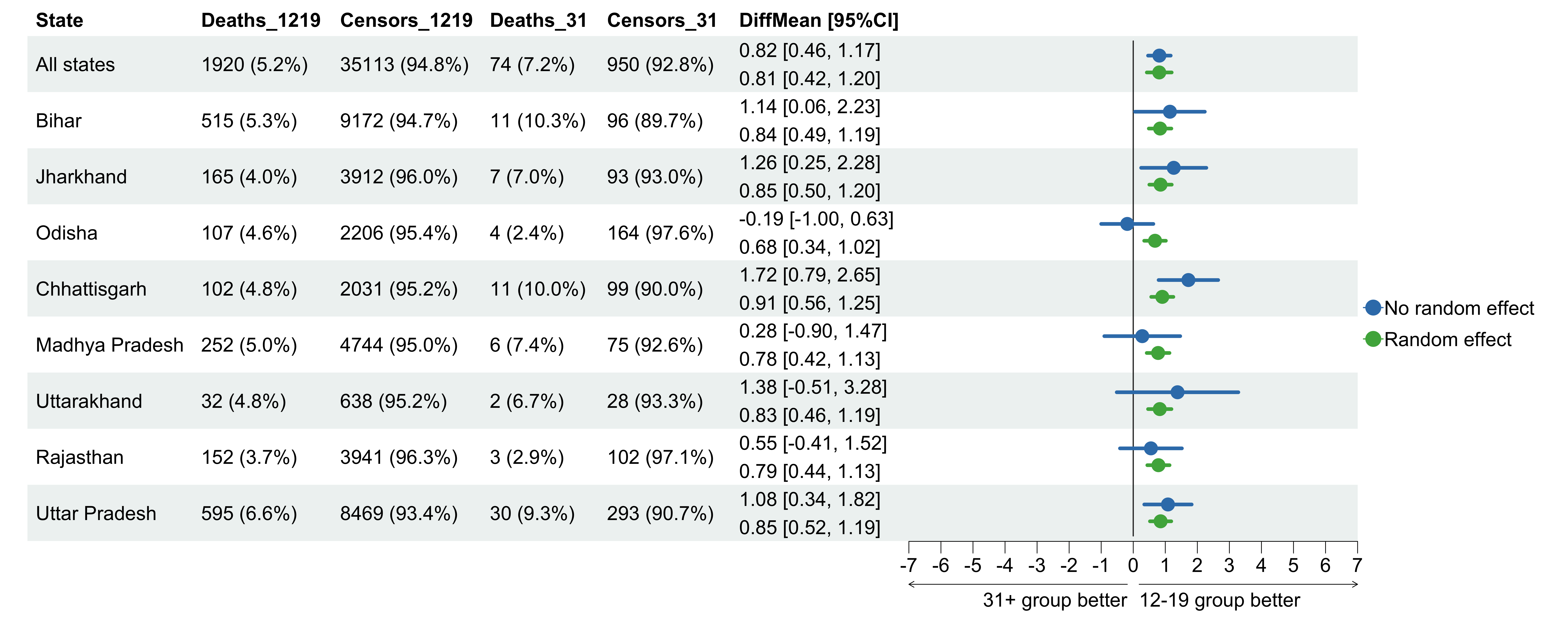}
  \caption{(PV method) Forest plot of the difference between the restricted mean survival times (12--19 group vs 31+ group)}
  \label{Andersen2}
        \footnotesize{}
  \end{center}
\end{figure}

\begin{figure}[H]
  \begin{center}
  \includegraphics[width=15cm]{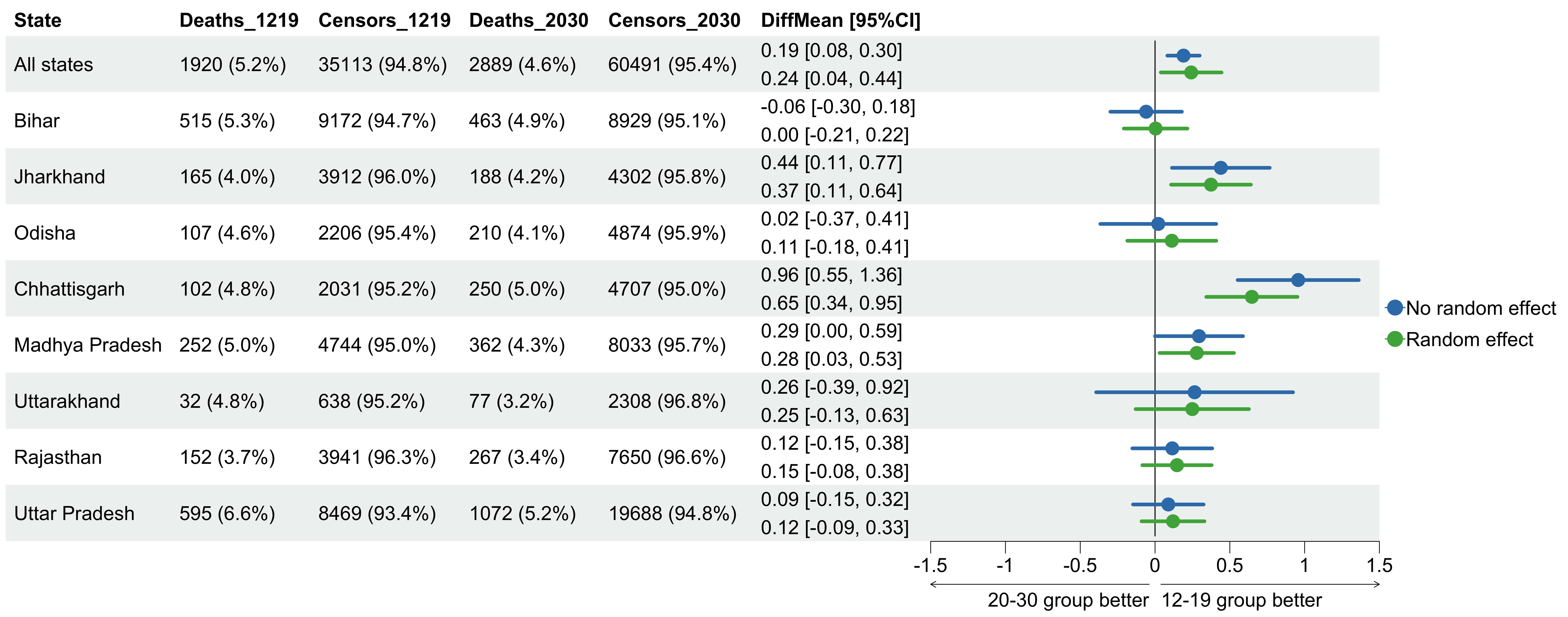}
  \caption{(IPCW method) Forest plot of the difference between the restricted mean survival times (12--19 group vs 20--30 group)}
  \label{Tian1}
        \footnotesize{}
  \end{center}
\end{figure}

\begin{figure}[H]
  \begin{center}
  \includegraphics[width=15cm]{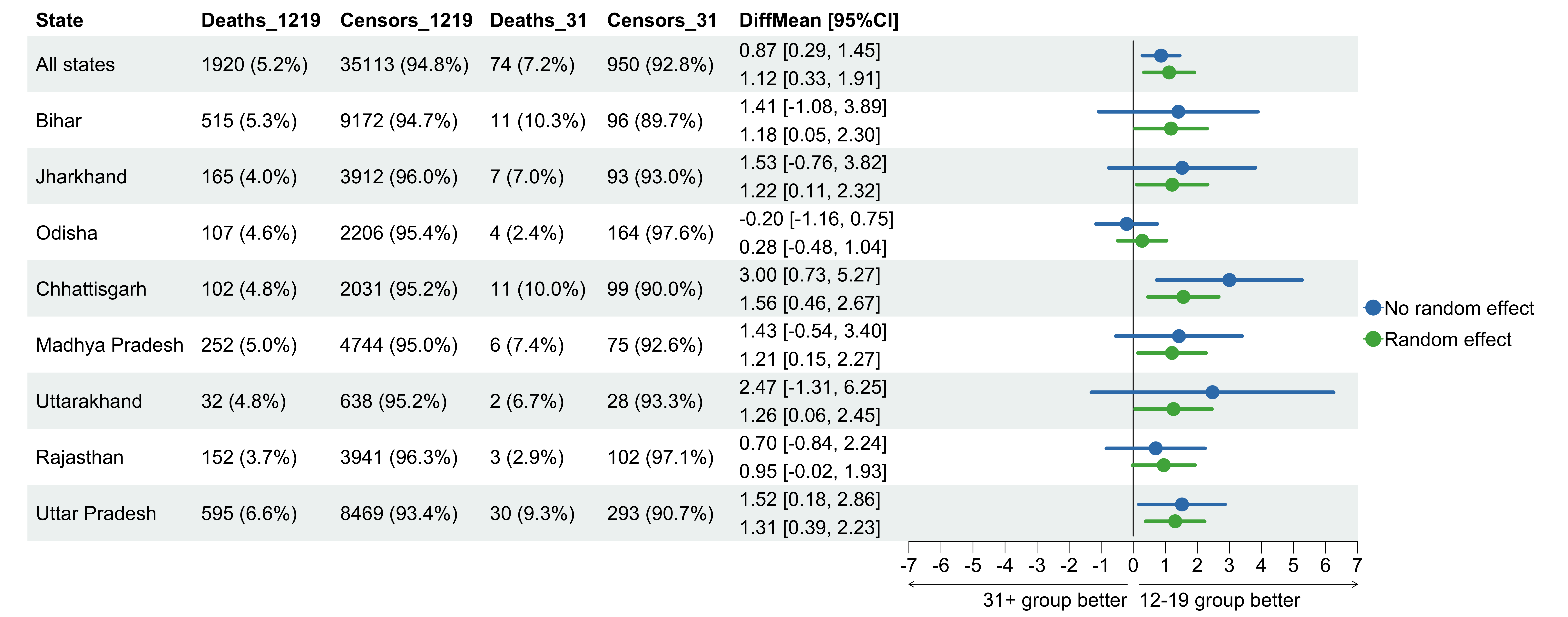}
  \caption{(IPCW method) Forest plot of the difference between the restricted mean survival times (12--19 group vs 31+ group)}
  \label{Tian2}
        \footnotesize{}
  \end{center}
\end{figure}

We re-analyzed the survival data of children up to age 5 years in India during the years 2019--2021. In India, eight EAG states were established to facilitate regional programs in areas that have lagged behind in controlling population growth to manageable levels. We focused on the survival data of the EAG states. The total sample size is 101,437, which consists of 37,033 individuals born to mothers 12--19 years old (12--19 group), 63,380 individuals born to mothers 20--30 years old (20--30 group), and 1,024 individuals born to mothers 31 years and older (31+ group). The total number of deaths is 4,883, comprising 1,920, 2,889, and 74 deaths for those born to mothers in the 12--19, 20--30, and 31+ groups, respectively. Moreover, the censoring probabilities are 94.8\%, 95.4\%, and 92.8\%, respectively. The baseline information consists of the sex, place of birth, and order of birth.
The number of deaths and censors for each state are shown on the left side in Figures \ref{Andersen1}, \ref{Andersen2}, \ref{Tian1}, and \ref{Tian2}.
We also re-analyzed data of 2019--2021 of an Indian national family health survey that included data for children up to age 5 years. The restricted time, which is the time at which we can identify more than 80\% of the deaths (i.e., 4,094 out of 4,883 total deaths), is 50 months.

The estimated random effects $\sih_v^2$ of the PV method are 0.026 for 12--19 group vs 20--30 group and 0.036 for 12--19 group vs 31+ group. The estimated random effects $\sih_v^2$ of IPCW method are 0.056 for 12--19 group vs 20--30 group and 0.414 for 12--19 group vs 31+ group. The RMST differences are shown in Figures \ref{Andersen1}, \ref{Andersen2}, \ref{Tian1}, and \ref{Tian2}. 

For the PV method, when random effects are accounted for, the results comparing the 20--30 group with the 12--19 group are reduced with respect to the overall results because of the small variance among the states. In particular, in Chhattisgarh, where the RMST difference deviates significantly from the overall result, the reduction was larger because the small sample size was found to bias the results by chance. For the 31+ group versus the 12--19 group, because of the small variance among states and the much smaller sample size for the 12--19 group, the results for each state obtained by the random-effect RMST are substantially reduced. This means that in all states, the prognosis is better if the child is born to a mother who is 12--19 years.

For the IPCW method, the results comparing the 20--30 group with the 12--19 group reveal that, as for the PV method, the variance among states is small; hence, results that deviate from the overall results, such as the result for Chhattisgarh, are reduced with respect to the overall results. In addition, when the 95\% confidence intervals are too long because of the small sample size, as in the result for Uttarakhand, the 95\% confidence intervals are shorter. For the 12--19 group vs 31+ group, because the IPCW method found a greater variance among the states than the PV method, the 95\% confidence interval was not shortened as much as it was by the PV method, but the 95\% confidence interval could be shortened for results from states with long confidence intervals.

\section{Discussion}
We proposed two novel random-effect RMST models. One imputes PVs for censored data to create a complete dataset and use a generalized linear mixed model. The other estimates the variance parameter of the random effect among clusters using the estimation results obtained by the IPCW estimating equation for each cluster. The real data analysis revealed that the shrinkage effect using the PV method is stronger than that obtained using the IPCW method because the number of interest events increases when the PVs of censored data are imputed.

The random-effect RMST model using the PV method analyzes all data by imputing the PVs for censored data. Given a large sample size, the individual PVs are approximately independent~\cite{andersen2003generalised}, and the estimators obtained from the generalized linear mixed model are consistent~\cite{jiang1998consistent,jiang2013subset}.
However, a limitation is that the PVs are not imputed while taking into account the random effects. For the IPCW method, because an estimated equation cannot be constructed that considers the random effects, we analyzed within each cluster and estimated the random effects of the clusters. The estimator is consistent with respect to the true parameter value~\cite{yoshimori2014second,slud2006mean}. The limitation of this approach is that it is not possible to perform an analysis that adjusts for random effects and covariates using data from all clusters.

The simulation study showed that as the sample size increased, the MSE decreased and the estimation accuracy improved. For the PV method, as the censoring probability increased, the accuracy of the estimated coefficient parameters did not change, but the results of the estimated parameters of the variance of the random effects decreased. The accuracy of the estimated random effect increased because of the smaller observation error for the PVs for censored data. Moreover, the results remained stable despite the increased censoring probability because the PVs were working effectively. For the IPCW method, the accuracy of the coefficient parameter estimates decreased as the censoring probability increased because the number of observed events was reduced. In addition, the estimated results for the parameter of variance of the random effect grew worse as the number of clusters increased. However, increasing the sample size increased the accuracy of estimation. With both methods, the coverage probability approached the true value of 0.95 as the accuracy of the coefficient parameter estimation increased.

From the actual study, for the 12--19 group vs 31+ group comparison, when the number of cases was small, the variance between states differed in two ways. We believe that this result reflects the results of the simulated study, since the IPCW method is less accurate in estimation when the number of clusters is large and the sample size is small. The PV method, which stores the censoring as a PV, analyzes a larger number of data as time-to-event, and hence the confidence intervals were shorter than those of the IPCW method, regardless of the presence of random effects.

We recommend the random-effect RMST model based on the PV method because it exhibited less bias in the simulation results. This is because the 95\% confidence interval was closer to the 95\% probability of containing the true value.

\bibliography{main.bib}
\bibliographystyle{unsrt}

\newpage

\begin{table}[H]
\begin{center}
\caption{Simulation results of model 1.}
\label{table:model1}
\small
\begin{tabular}{|cccccc|}
\hline
        & \multicolumn{1}{c|}{Censoring}   & \multicolumn{2}{c|}{PV method}                  & \multicolumn{2}{c|}{IPCW method} \\
Cluster & \multicolumn{1}{c|}{probability} & Total N=400 & \multicolumn{1}{c|}{Total N=1000} & Total N=400    & Total N=1000    \\ \hline
\multicolumn{6}{|l|}{Bias ($\beta_1$)}                                                                                          \\ \hline
5       & \multicolumn{1}{c|}{0.1}         & -0.010      & \multicolumn{1}{c|}{-0.009}       & -0.008         & -0.008          \\
8       & \multicolumn{1}{c|}{0.1}         & -0.002      & \multicolumn{1}{c|}{-0.008}       & 0.004          & -0.005          \\
10      & \multicolumn{1}{c|}{0.1}         & 0.010       & \multicolumn{1}{c|}{-0.008}       & 0.019          & -0.005          \\
5       & \multicolumn{1}{c|}{0.5}         & -0.014      & \multicolumn{1}{c|}{-0.020}       & -0.003         & -0.028          \\
8       & \multicolumn{1}{c|}{0.5}         & -0.012      & \multicolumn{1}{c|}{-0.020}       & 0.019          & -0.024          \\
10      & \multicolumn{1}{c|}{0.5}         & -0.012      & \multicolumn{1}{c|}{-0.016}       & 0.038          & -0.015          \\ \hline
\multicolumn{6}{|l|}{MSE ($\beta_1$)}                                                                                           \\ \hline
5       & \multicolumn{1}{c|}{0.1}         & 0.034       & \multicolumn{1}{c|}{0.013}        & 0.036          & 0.014           \\
8       & \multicolumn{1}{c|}{0.1}         & 0.034       & \multicolumn{1}{c|}{0.013}        & 0.036          & 0.014           \\
10      & \multicolumn{1}{c|}{0.1}         & 0.037       & \multicolumn{1}{c|}{0.013}        & 0.040          & 0.014           \\
5       & \multicolumn{1}{c|}{0.5}         & 0.039       & \multicolumn{1}{c|}{0.016}        & 0.060          & 0.025           \\
8       & \multicolumn{1}{c|}{0.5}         & 0.039       & \multicolumn{1}{c|}{0.016}        & 0.064          & 0.024           \\
10      & \multicolumn{1}{c|}{0.5}         & 0.041       & \multicolumn{1}{c|}{0.015}        & 0.073          & 0.023           \\ \hline
\multicolumn{6}{|l|}{Coverage probability of the confidence interval of $\beta_1$}                                                                         \\ \hline
5       & \multicolumn{1}{c|}{0.1}         & 0.936       & \multicolumn{1}{c|}{0.940}        & 0.936          & 0.940           \\
8       & \multicolumn{1}{c|}{0.1}         & 0.936       & \multicolumn{1}{c|}{0.939}        & 0.928          & 0.938           \\
10      & \multicolumn{1}{c|}{0.1}         & 0.920       & \multicolumn{1}{c|}{0.940}        & 0.912          & 0.936           \\
5       & \multicolumn{1}{c|}{0.5}         & 0.932       & \multicolumn{1}{c|}{0.931}        & 0.885          & 0.879           \\
8       & \multicolumn{1}{c|}{0.5}         & 0.929       & \multicolumn{1}{c|}{0.934}        & 0.871          & 0.891           \\
10      & \multicolumn{1}{c|}{0.5}         & 0.924       & \multicolumn{1}{c|}{0.934}        & 0.838          & 0.890           \\ \hline
\multicolumn{6}{|l|}{Length of the confidence interval of $\beta_1$}                                                                \\ \hline
5       & \multicolumn{1}{c|}{0.1}         & 0.685       & \multicolumn{1}{c|}{0.436}        & 0.698          & 0.444           \\
8       & \multicolumn{1}{c|}{0.1}         & 0.679       & \multicolumn{1}{c|}{0.435}        & 0.690          & 0.443           \\
10      & \multicolumn{1}{c|}{0.1}         & 0.675       & \multicolumn{1}{c|}{0.434}        & 0.685          & 0.442           \\
5       & \multicolumn{1}{c|}{0.5}         & 0.725       & \multicolumn{1}{c|}{0.461}        & 0.777          & 0.498           \\
8       & \multicolumn{1}{c|}{0.5}         & 0.718       & \multicolumn{1}{c|}{0.459}        & 0.759          & 0.494           \\
10      & \multicolumn{1}{c|}{0.5}         & 0.712       & \multicolumn{1}{c|}{0.459}        & 0.745          & 0.492           \\ \hline
\multicolumn{6}{|l|}{Bias ($\sigma_v^2$)}                                                                                       \\ \hline
5       & \multicolumn{1}{c|}{0.1}         & 0.067       & \multicolumn{1}{c|}{0.074}        & -0.040         & -0.069          \\
8       & \multicolumn{1}{c|}{0.1}         & 0.066       & \multicolumn{1}{c|}{0.074}        & -0.018         & -0.064          \\
10      & \multicolumn{1}{c|}{0.1}         & 0.064       & \multicolumn{1}{c|}{0.072}        & -0.002         & -0.060          \\
5       & \multicolumn{1}{c|}{0.5}         & 0.003       & \multicolumn{1}{c|}{0.013}        & 0.018          & -0.039          \\
8       & \multicolumn{1}{c|}{0.5}         & 0.001       & \multicolumn{1}{c|}{0.012}        & 0.085          & -0.026          \\
10      & \multicolumn{1}{c|}{0.5}         & 0.001       & \multicolumn{1}{c|}{0.010}        & 0.155          & -0.017          \\ \hline
\multicolumn{6}{|l|}{MSE ($\sigma_v^2$)}                                                                                        \\ \hline
5       & \multicolumn{1}{c|}{0.1}         & 0.025       & \multicolumn{1}{c|}{0.021}        & 0.010          & 0.006           \\
8       & \multicolumn{1}{c|}{0.1}         & 0.019       & \multicolumn{1}{c|}{0.016}        & 0.014          & 0.006           \\
10      & \multicolumn{1}{c|}{0.1}         & 0.018       & \multicolumn{1}{c|}{0.014}        & 0.018          & 0.006           \\
5       & \multicolumn{1}{c|}{0.5}         & 0.013       & \multicolumn{1}{c|}{0.009}        & 0.031          & 0.007           \\
8       & \multicolumn{1}{c|}{0.5}         & 0.011       & \multicolumn{1}{c|}{0.006}        & 0.067          & 0.008           \\
10      & \multicolumn{1}{c|}{0.5}         & 0.010       & \multicolumn{1}{c|}{0.005}        & 0.125          & 0.009           \\ \hline
\end{tabular}
\end{center}
\end{table}

\begin{table}[H]
\begin{center}
\caption{Simulation results of model 2.}
\label{table:model2}
\small
\begin{tabular}{|cccccc|}
\hline
        & \multicolumn{1}{c|}{Censoring}   & \multicolumn{2}{c|}{PV method}                  & \multicolumn{2}{c|}{IPCW method} \\
Cluster & \multicolumn{1}{c|}{probability} & Total N=400 & \multicolumn{1}{c|}{Total N=1000} & Total N=400    & Total N=1000    \\ \hline
\multicolumn{6}{|l|}{Bias ($\beta_1$)}                                                                                          \\ \hline
5       & \multicolumn{1}{c|}{0.1}         & -0.015      & \multicolumn{1}{c|}{-0.020}       & -0.012         & -0.020          \\
8       & \multicolumn{1}{c|}{0.1}         & -0.009      & \multicolumn{1}{c|}{-0.017}       & -0.003         & -0.015          \\
10      & \multicolumn{1}{c|}{0.1}         & -0.003      & \multicolumn{1}{c|}{-0.016}       & 0.004          & -0.013          \\
5       & \multicolumn{1}{c|}{0.5}         & 0.005       & \multicolumn{1}{c|}{0.005}        & 0.031          & 0.006           \\
8       & \multicolumn{1}{c|}{0.5}         & 0.010       & \multicolumn{1}{c|}{0.004}        & 0.062          & 0.011           \\
10      & \multicolumn{1}{c|}{0.5}         & 0.018       & \multicolumn{1}{c|}{0.006}        & 0.097          & 0.021           \\ \hline
\multicolumn{6}{|l|}{MSE ($\beta_1$)}                                                                                           \\ \hline
5       & \multicolumn{1}{c|}{0.1}         & 0.033       & \multicolumn{1}{c|}{0.013}        & 0.035          & 0.014           \\
8       & \multicolumn{1}{c|}{0.1}         & 0.034       & \multicolumn{1}{c|}{0.013}        & 0.036          & 0.014           \\
10      & \multicolumn{1}{c|}{0.1}         & 0.035       & \multicolumn{1}{c|}{0.013}        & 0.038          & 0.014           \\
5       & \multicolumn{1}{c|}{0.5}         & 0.042       & \multicolumn{1}{c|}{0.017}        & 0.069          & 0.027           \\
8       & \multicolumn{1}{c|}{0.5}         & 0.042       & \multicolumn{1}{c|}{0.016}        & 0.079          & 0.026           \\
10      & \multicolumn{1}{c|}{0.5}         & 0.044       & \multicolumn{1}{c|}{0.016}        & 0.095          & 0.026           \\ \hline
\multicolumn{6}{|l|}{Coverage probability of the confidence interval of $\beta_1$}                                                                         \\ \hline
5       & \multicolumn{1}{c|}{0.1}         & 0.941       & \multicolumn{1}{c|}{0.941}        & 0.936          & 0.939           \\
8       & \multicolumn{1}{c|}{0.1}         & 0.934       & \multicolumn{1}{c|}{0.942}        & 0.931          & 0.941           \\
10      & \multicolumn{1}{c|}{0.1}         & 0.924       & \multicolumn{1}{c|}{0.942}        & 0.919          & 0.942           \\
5       & \multicolumn{1}{c|}{0.5}         & 0.935       & \multicolumn{1}{c|}{0.938}        & 0.882          & 0.894           \\
8       & \multicolumn{1}{c|}{0.5}         & 0.929       & \multicolumn{1}{c|}{0.938}        & 0.850          & 0.899           \\
10      & \multicolumn{1}{c|}{0.5}         & 0.922       & \multicolumn{1}{c|}{0.941}        & 0.802          & 0.897           \\ \hline
\multicolumn{6}{|l|}{Length of the confidence interval of $\beta_1$}                                                                \\ \hline
5       & \multicolumn{1}{c|}{0.1}         & 0.683       & \multicolumn{1}{c|}{0.434}        & 0.697          & 0.444           \\
8       & \multicolumn{1}{c|}{0.1}         & 0.677       & \multicolumn{1}{c|}{0.433}        & 0.690          & 0.443           \\
10      & \multicolumn{1}{c|}{0.1}         & 0.672       & \multicolumn{1}{c|}{0.432}        & 0.684          & 0.442           \\
5       & \multicolumn{1}{c|}{0.5}         & 0.759       & \multicolumn{1}{c|}{0.482}        & 0.825          & 0.530           \\
8       & \multicolumn{1}{c|}{0.5}         & 0.753       & \multicolumn{1}{c|}{0.481}        & 0.804          & 0.526           \\
10      & \multicolumn{1}{c|}{0.5}         & 0.748       & \multicolumn{1}{c|}{0.480}        & 0.787          & 0.523           \\ \hline
\multicolumn{6}{|l|}{Bias ($\sigma_v^2$)}                                                                                       \\ \hline
5       & \multicolumn{1}{c|}{0.1}         & 0.065       & \multicolumn{1}{c|}{0.072}        & -0.038         & -0.071          \\
8       & \multicolumn{1}{c|}{0.1}         & 0.064       & \multicolumn{1}{c|}{0.072}        & -0.019         & -0.065          \\
10      & \multicolumn{1}{c|}{0.1}         & 0.063       & \multicolumn{1}{c|}{0.072}        & -0.002         & -0.060          \\
5       & \multicolumn{1}{c|}{0.5}         & 0.012       & \multicolumn{1}{c|}{0.021}        & 0.032          & -0.040          \\
8       & \multicolumn{1}{c|}{0.5}         & 0.008       & \multicolumn{1}{c|}{0.019}        & 0.111          & -0.023          \\
10      & \multicolumn{1}{c|}{0.5}         & 0.009       & \multicolumn{1}{c|}{0.019}        & 0.195          & -0.011          \\ \hline
\multicolumn{6}{|l|}{MSE ($\sigma_v^2$)}                                                                                        \\ \hline
5       & \multicolumn{1}{c|}{0.1}         & 0.024       & \multicolumn{1}{c|}{0.021}        & 0.010          & 0.006           \\
8       & \multicolumn{1}{c|}{0.1}         & 0.019       & \multicolumn{1}{c|}{0.015}        & 0.014          & 0.006           \\
10      & \multicolumn{1}{c|}{0.1}         & 0.017       & \multicolumn{1}{c|}{0.013}        & 0.018          & 0.006           \\
5       & \multicolumn{1}{c|}{0.5}         & 0.015       & \multicolumn{1}{c|}{0.010}        & 0.039          & 0.008           \\
8       & \multicolumn{1}{c|}{0.5}         & 0.012       & \multicolumn{1}{c|}{0.007}        & 0.092          & 0.009           \\
10      & \multicolumn{1}{c|}{0.5}         & 0.012       & \multicolumn{1}{c|}{0.006}        & 0.167          & 0.011           \\ \hline
\end{tabular}
\end{center}
\end{table}

\appendix
\section{Appendix}\label{App}

\subsection{Simulated results for a censoring probability of 0.9}
\label{ap:cnsr0.9}

\begin{table}[H]
\begin{center}
\caption{Simulation results for a censoring probability of 0.9.}
\label{table:cnsr0.9}
\scalebox{0.8}{
\begin{tabular}{|cccccc|}
\hline
        & \multicolumn{1}{c|}{}      & \multicolumn{2}{c|}{PV method}                  & \multicolumn{2}{c|}{IPCW method} \\
Cluster & \multicolumn{1}{c|}{Model} & Total N=400 & \multicolumn{1}{c|}{Total N=1000} & Total N=400    & Total N=1000    \\ \hline
\multicolumn{6}{|l|}{Bias ($\beta_1$)}                                                                                    \\ \hline
5       & \multicolumn{1}{c|}{1}     & -0.018      & \multicolumn{1}{c|}{-0.006}       & 0.023          & -0.008          \\
8       & \multicolumn{1}{c|}{1}     & -0.052      & \multicolumn{1}{c|}{-0.013}       & 0.056          & -0.010          \\
10      & \multicolumn{1}{c|}{1}     & -0.082      & \multicolumn{1}{c|}{-0.019}       & 0.039          & -0.008          \\
5       & \multicolumn{1}{c|}{2}     & -0.019      & \multicolumn{1}{c|}{-0.005}       & 0.049          & -0.008          \\
8       & \multicolumn{1}{c|}{2}     & -0.050      & \multicolumn{1}{c|}{-0.014}       & 0.135          & -0.001          \\
10      & \multicolumn{1}{c|}{2}     & -0.082      & \multicolumn{1}{c|}{-0.018}       & 0.152          & 0.013           \\ \hline
\multicolumn{6}{|l|}{MSE ($\beta_1$)}                                                                                     \\ \hline
5       & \multicolumn{1}{c|}{1}     & 0.015       & \multicolumn{1}{c|}{0.006}        & 0.098          & 0.016           \\
8       & \multicolumn{1}{c|}{1}     & 0.016       & \multicolumn{1}{c|}{0.006}        & 0.322          & 0.022           \\
10      & \multicolumn{1}{c|}{1}     & 0.018       & \multicolumn{1}{c|}{0.006}        & 0.431          & 0.029           \\
5       & \multicolumn{1}{c|}{2}     & 0.018       & \multicolumn{1}{c|}{0.008}        & 0.151          & 0.022           \\
8       & \multicolumn{1}{c|}{2}     & 0.019       & \multicolumn{1}{c|}{0.007}        & 0.463          & 0.036           \\
10      & \multicolumn{1}{c|}{2}     & 0.022       & \multicolumn{1}{c|}{0.008}        & 0.779          & 0.056           \\ \hline
\multicolumn{6}{|l|}{Coverage probability of the confidence interval of $\beta_1$}                                                                   \\ \hline
5       & \multicolumn{1}{c|}{1}     & 0.890       & \multicolumn{1}{c|}{0.926}        & 0.717          & 0.797           \\
8       & \multicolumn{1}{c|}{1}     & 0.713       & \multicolumn{1}{c|}{0.918}        & 0.480          & 0.758           \\
10      & \multicolumn{1}{c|}{1}     & 0.502       & \multicolumn{1}{c|}{0.897}        & 0.316          & 0.708           \\
5       & \multicolumn{1}{c|}{2}     & 0.905       & \multicolumn{1}{c|}{0.931}        & 0.686          & 0.780           \\
8       & \multicolumn{1}{c|}{2}     & 0.758       & \multicolumn{1}{c|}{0.923}        & 0.453          & 0.722           \\
10      & \multicolumn{1}{c|}{2}     & 0.563       & \multicolumn{1}{c|}{0.910}        & 0.307          & 0.680           \\ \hline
\multicolumn{6}{|l|}{Length of the confidence interval of $\beta_1$}                                                          \\ \hline
5       & \multicolumn{1}{c|}{1}     & 0.423       & \multicolumn{1}{c|}{0.288}        & 0.437          & 0.304           \\
8       & \multicolumn{1}{c|}{1}     & 0.330       & \multicolumn{1}{c|}{0.278}        & 0.330          & 0.290           \\
10      & \multicolumn{1}{c|}{1}     & 0.230       & \multicolumn{1}{c|}{0.270}        & 0.241          & 0.280           \\
5       & \multicolumn{1}{c|}{2}     & 0.474       & \multicolumn{1}{c|}{0.319}        & 0.495          & 0.342           \\
8       & \multicolumn{1}{c|}{2}     & 0.385       & \multicolumn{1}{c|}{0.309}        & 0.383          & 0.327           \\
10      & \multicolumn{1}{c|}{2}     & 0.283       & \multicolumn{1}{c|}{0.301}        & 0.291          & 0.316           \\ \hline
\multicolumn{6}{|l|}{Bias ($\sigma_v^2$)}                                                                                 \\ \hline
5       & \multicolumn{1}{c|}{1}     & -0.087      & \multicolumn{1}{c|}{-0.083}       & 0.156          & -0.032          \\
8       & \multicolumn{1}{c|}{1}     & -0.089      & \multicolumn{1}{c|}{-0.084}       & 0.462          & 0.015           \\
10      & \multicolumn{1}{c|}{1}     & -0.088      & \multicolumn{1}{c|}{-0.084}       & 0.595          & 0.063           \\
5       & \multicolumn{1}{c|}{2}     & -0.086      & \multicolumn{1}{c|}{-0.082}       & 0.291          & -0.012          \\
8       & \multicolumn{1}{c|}{2}     & -0.088      & \multicolumn{1}{c|}{-0.083}       & 0.723          & 0.081           \\
10      & \multicolumn{1}{c|}{2}     & -0.086      & \multicolumn{1}{c|}{-0.083}       & 0.952          & 0.166           \\ \hline
\multicolumn{6}{|l|}{MSE ($\sigma_v^2$)}                                                                                  \\ \hline
5       & \multicolumn{1}{c|}{1}     & 0.008       & \multicolumn{1}{c|}{0.007}        & 0.483          & 0.026           \\
8       & \multicolumn{1}{c|}{1}     & 0.009       & \multicolumn{1}{c|}{0.007}        & 1.798          & 0.079           \\
10      & \multicolumn{1}{c|}{1}     & 0.009       & \multicolumn{1}{c|}{0.007}        & 2.726          & 0.144           \\
5       & \multicolumn{1}{c|}{2}     & 0.009       & \multicolumn{1}{c|}{0.007}        & 0.907          & 0.043           \\
8       & \multicolumn{1}{c|}{2}     & 0.009       & \multicolumn{1}{c|}{0.007}        & 2.557          & 0.173           \\
10      & \multicolumn{1}{c|}{2}     & 0.010       & \multicolumn{1}{c|}{0.007}        & 4.680          & 0.327           \\ \hline
\end{tabular}
}
\end{center}
\end{table}

\subsection{Simulated true value of parameter $\beta_1$}
\label{ap:sim_beta}

\begin{table}[H]
\centering
\caption{Simulated true value of parameter $\bar{\beta}_1$}
\label{table:ex_rmst}
\begin{tabular}{|c|c|cc|c|}
\hline
Model                  & Method                & \begin{tabular}[c]{@{}c@{}}Censoring\\ probability\end{tabular} & \begin{tabular}[c]{@{}c@{}}Number of\\ simulations\end{tabular} & $\bar{\beta}_1$ \\ \hline
Model 1                & PV              & 0.1                                                             & 100000                                                    & 0.7156          \\
                       &                       & 0.5                                                             & 100000                                                    & 0.5681          \\
\multicolumn{1}{|l|}{} & \multicolumn{1}{l|}{} & 0.9                                                             & 100000                                                    & 0.1531          \\ \cline{2-5} 
                       & IPCW                  & 0.1                                                             & 100000                                                    & 0.7623          \\
                       &                       & 0.5                                                             & 100000                                                    & 0.8086          \\
\multicolumn{1}{|l|}{} & \multicolumn{1}{l|}{} & 0.9                                                             & 100000                                                    & 0.3057          \\ \hline
Model 2                & PV              & 0.1                                                             & 100000                                                    & 0.7156          \\
                       &                       & 0.5                                                             & 100000                                                    & 0.5681          \\
\multicolumn{1}{|l|}{} & \multicolumn{1}{l|}{} & 0.9                                                             & 100000                                                    & 0.1682          \\ \cline{2-5} 
                       & IPCW                  & 0.1                                                             & 100000                                                    & 0.7623          \\
                       &                       & 0.5                                                             & 100000                                                    & 0.8086          \\
\multicolumn{1}{|l|}{} & \multicolumn{1}{l|}{} & 0.9                                                             & 100000                                                    & 0.3566          \\ \hline
\end{tabular}
\end{table}

\end{document}